\documentclass[conf, onecolumn]{ao4elt3}
\usepackage[]{graphicx,amsmath,xspace} 
\usepackage[caption=false]{subfig}
\usepackage[varg]{txfonts}
\usepackage[latin1]{inputenc}

\newcommand{\FIG}[3]{\includegraphics[width=#1\linewidth,draft=#2]{#3.eps}}

\newcommand{\loD}{\mbox{$\lambda/D$}\xspace}
\newcommand{\e}[1]{10^{#1}}
\newcommand{\E}[1]{\times10^{#1}}
\newcommand{\Coro}{Coronagraph\xspace}

\newcommand{\coro}{coronagraph\xspace}
\newcommand{\coros}{coronagraphs\xspace}

\graphicspath{{D:/Documents/2_EXCEDE/DATA/}{D:/Documents/2_EXCEDE/REPORTS/}}

\begin{document}

\pagestyle{plain}

\title{\vspace{1.6cm}\\EXPERIMENTAL STUDY OF A LOW-ORDER WAVEFRONT SENSOR FOR A HIGH-CONTRAST CORONAGRAPHIC IMAGER AT 1.2~\loD}

\author{Julien Lozi\inst{1}\thanks{jlozi@email.arizona.edu} \and Ruslan~Belikov\inst{2} \and Glenn~Schneider\inst{1} \and Olivier~Guyon\inst{1} \and Eugene~Pluzhnik\inst{2,3} \and Sandrine~J.~Thomas\inst{2,3} \and Frantz~Martinache\inst{4}}

\institute{University of Arizona, 1401 E University Blvd, Tucson, AZ 85721, USA \and NASA Ames Research Center, Moffett Field, CA 94035, USA \and UARC/NASA Ames, P.O. Box 7, Moffett Field, CA 94035, USA \and Subaru Telescope, National Astronomical Observatory of Japan, 650 North A'ohoku Place, Hilo, HI 96720, USA}

\abstract{High-contrast imaging will be a challenge for future ELTs, because their vibrations create low-order aberrations ---~mostly tip/tilt~--- that reduce coronagraphic performances at 1.2~\loD and above. A Low-Order WaveFront Sensor (LOWFS) is essential to measure and control those aberrations. An experiment simulating a starlight suppression system is currently developed at NASA Ames Research Center, and includes a LOWFS controlling tip/tilt modes in real-time at 500~Hz. The LOWFS allowed us to reduce the tip/tilt disturbances to $\e{-3}$~\loD rms, enhancing the previous contrast by a decade, to $8\E{-7}$ between 1.2 and 2~\loD. A Linear Quadratic Gaussian (LQG) controller is currently implemented to improve even more that result by reducing residual vibrations. This testbed is developed for the mission EXCEDE (EXoplanetary Circumstellar Environments and Disk Explorer), selected by NASA for technology development, and designed to study the formation, evolution and architectures of exoplanetary systems and characterize circumstellar environments into stellar habitable zones. It is composed of a 0.7~m telescope equipped with a Phase-Induced Amplitude Apodization Coronagraph (PIAA-C) and a 2000-element MEMS deformable mirror, capable of raw contrasts of $\e{-6}$ at 1.2~\loD and $\e{-7}$ above 2~\loD. Although the testbed simulates space conditions, its LOWFS has the same design than on the SCExAO instrument at Subaru telescope, with whom it shares the same kind of problematic. Experimental results show that a good knowledge of the low-order disturbances is a key asset for high contrast imaging, whether for real-time control or for post processing, both in space and on ground telescopes.}
\maketitle
%

\section{INTRODUCTION}
\label{sec:intro}

In the next generation of ground and space \coros, stability of the instrument is a key issue to obtain good contrasts at small Inner Working Angles (IWA). Space telescopes are affected by vibrations, due to reaction wheels for example, as well as pointing stability, while ground telescopes are mostly affected by turbulence and vibrations. Therefore, if we want to image planets and stellar environments at 1~AU, \coros will have to be equipped with Low-Order Wavefront Sensors (LOWFS), capable of measuring and correcting low-order aberrations.

In the context of the EXCEDE (EXoplanetary Circumstellar Environments and Disk Explorer) mission \cite{Guyon12}, we are currently testing its starlight suppression system (the \coro with the wavefront correction and the LOWFS) in air, at the Ames Coronagraphic Experiment (ACE) at NASA Ames Research Center (Moffett Field, CA) and soon in vacuum at Lockheed Martin (Palo Alto, CA). A description of the goals of EXCEDE and ACE is presented in \cite{Belikov13}, while \cite{Thomas13b} presents the setup and the results of the wavefront control.

In this paper, we will present the experimental analysis and the results we obtained on the LOWFS we implemented in the ACE laboratory. Section~\ref{sec:TheLowOrderWavefrontSensor} provides a description of the principle of the LOWFS, as well as the hardware and software architecture. Then Sec.~\ref{sec:ExperimentalResults} presents the performances of the LOWFS and analyzes its sensitivity to different parameters. Finally, Sec.~\ref{sec:ApplicationToTheELTs} presents some conclusions about its value for ELTs.


\section{The Low-Order Wavefront Sensor}
\label{sec:TheLowOrderWavefrontSensor}


\subsection{Principle}
\label{sec:Principle}

The coronagraphic low-order wavefront sensor, developed by O.Guyon, is well described in \cite{Guyon09}. It uses a defocused image of the light rejected by the focal plane occulter. Images are then compared to a reference, and decomposed on a base of orthogonal modes. It was developed and successfully tested for PIAA \coros \cite{Guyon09,Vogt10}, especially the SCExAO instrument of the Subaru Telescope \cite{Martinache09}. This type of LOWFS is not exclusively designed for PIAA \coros, it can also be implemented on any type of \coro that uses a focal plane occulter.

A similar design was also developed for \coros with complex masks using the light rejected by the Lyot stop, such as PIAA Complex Mask \Coro (PIAACMC) \cite{Guyon10}, the vortex or the four quadrant phase mask (FQPM). It has been already tested on sky on SCExAO \cite{Singh13}. The real-time measurements can also be used during post-processing, to remove a maximum of residual starlight. This technique was successfully tested in a lab experiment \cite{Vogt11}.

For a PIAA \coro, the main modes we have to correct are the pre-PIAA tip and tilt. Because of the shape of the PIAA optics, pre-PIAA tip/tilt is different from post-PIAA tip/tilt. In a first order, the first one can compensate the second one, but if the disturbances are too large, both has to be controlled.


\subsection{Setup}
\label{sec:Setup}

The LOWFS is composed of three elements: first a 3-zone mask as described in \cite{Guyon09}: a transmissive zone for the planet light, an opaque disk to block the core of the star, and a reflective annulus to keep the edges of the star. Then a lens collects the light from the star reflected by the focal plane mask, and sends it to a camera. This lens does not need to be of very good quality or achromatic. Finally a fast camera that takes images of the starlight with a frequency high enough to correct eventual disturbances. In our experiment, we use a window of $50\times50$~pixels, and a frame rate of 1.1~kHz.

The images are grabbed by a real-time computer, a National Instrument PXI. It is used to perform fast calculations without jitters. Commands are then calculated and sent to actuators. In our case, we control the $x$- and $y$-axis translations of the source, corresponding to pre-PIAA tip/tilt, but the system is very flexible, and we can add other modes: focus, post-PIAA tip/tilt, astigmatism, etc.

The calibration of the LOWFS is done in two steps: We start by calibrating the reference position, then we calculate the modes we want to correct.

\begin{figure} \centering
  \subfloat[Reference image after a calibration.]
	{\label{fig:refim}\FIG{0.235}{false}{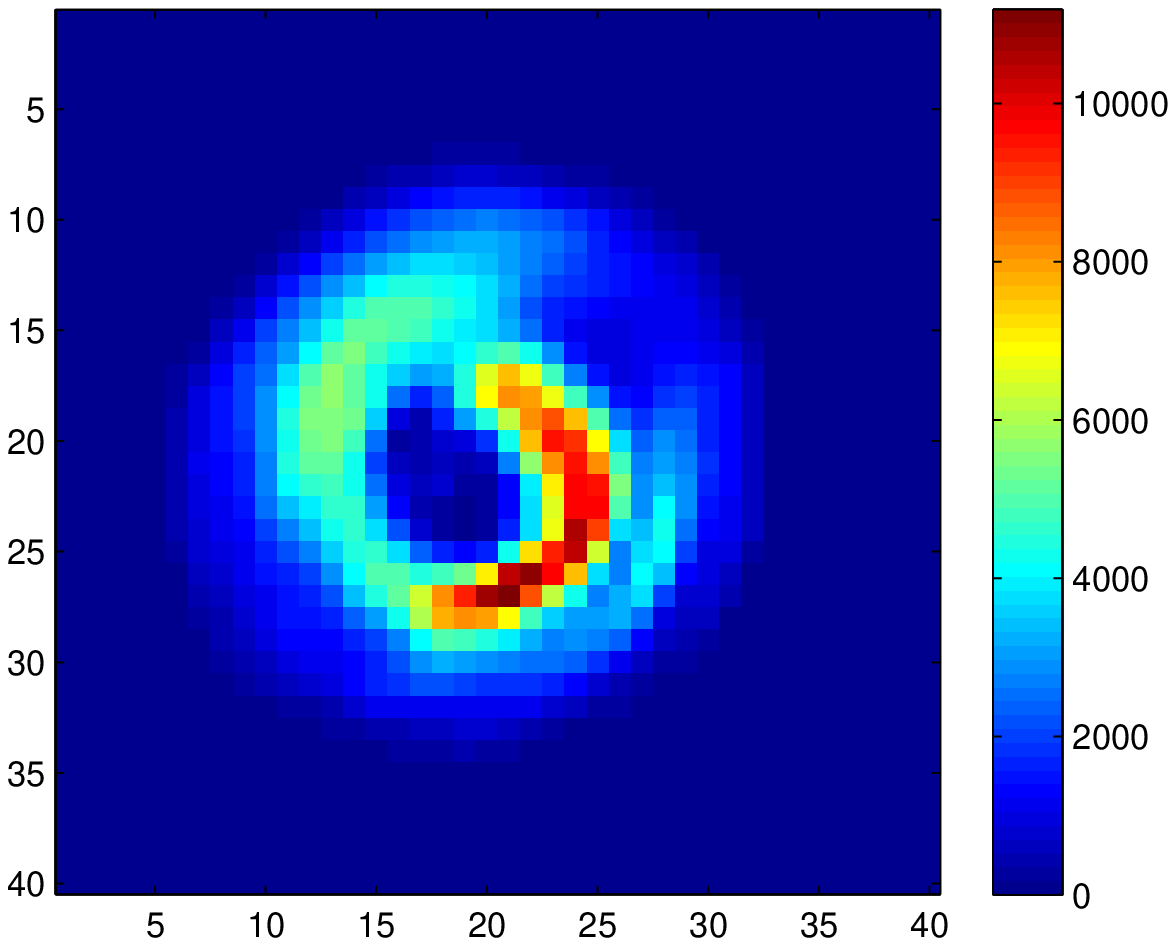}}\hspace{5pt}
  \subfloat[Difference b/w an image and the reference.]
	{\label{fig:signal}\FIG{0.235}{false}{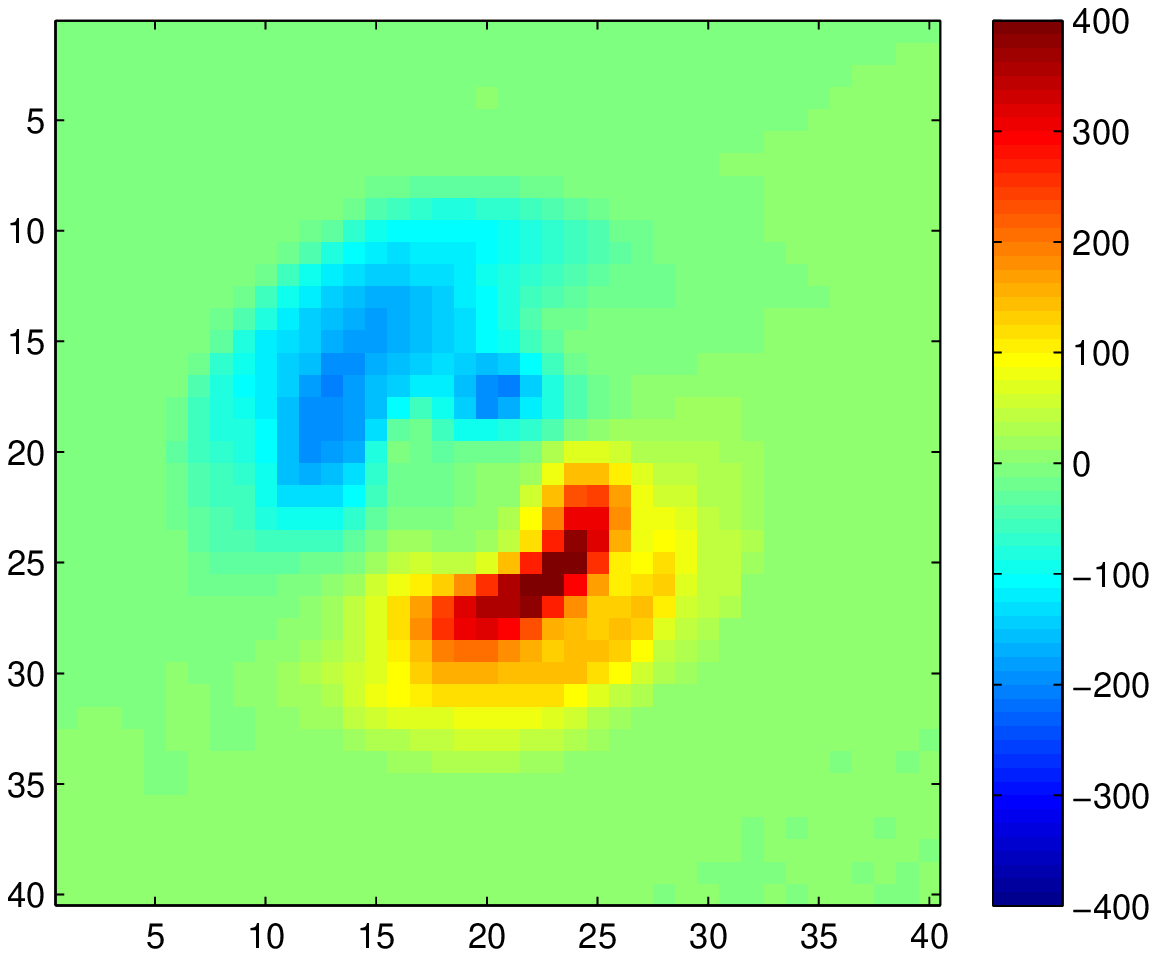}}\hspace{5pt}
  \subfloat[Mode 1 (pre-PIAA tip) of the LOWFS.]
	{\label{fig:mode1}\FIG{0.235}{false}{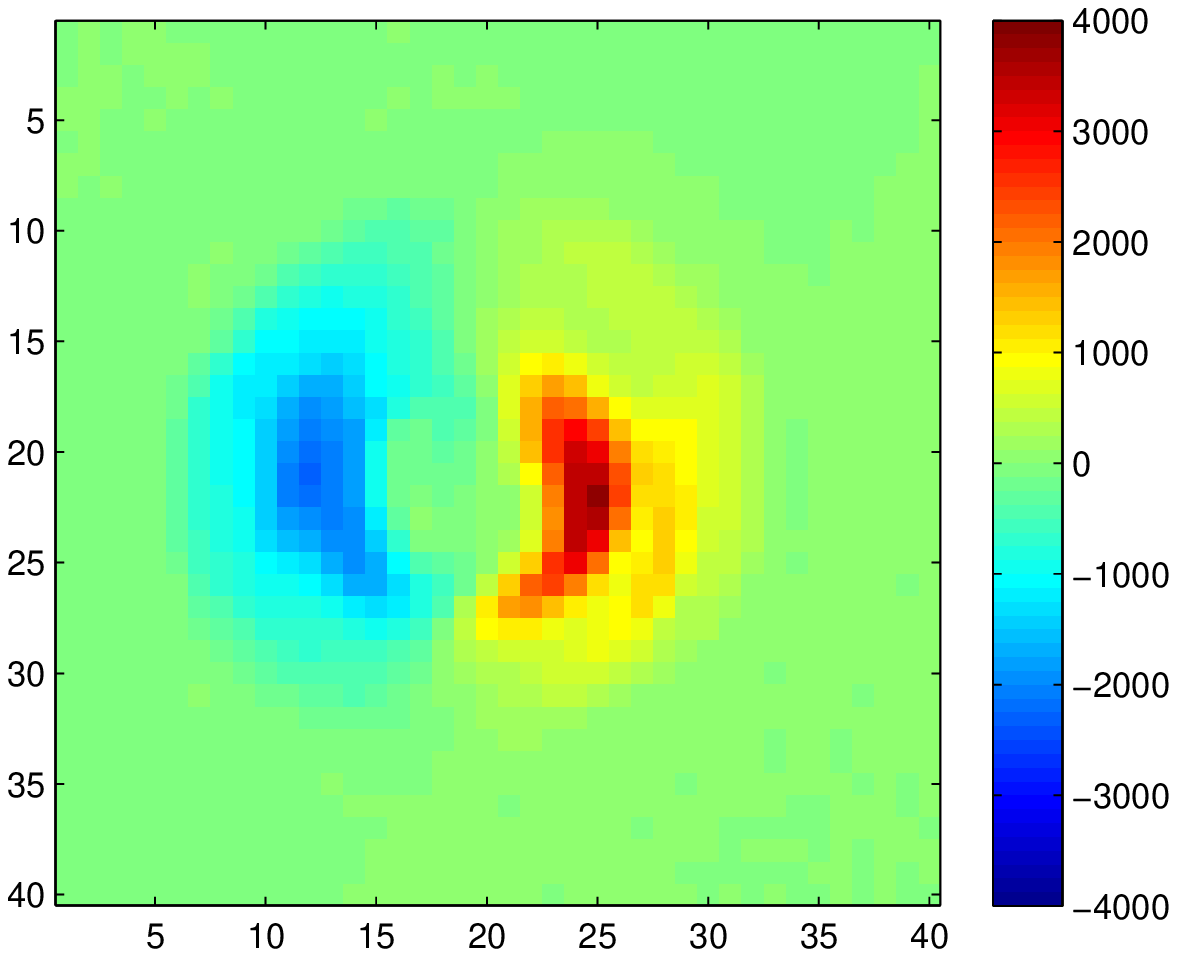}}\hspace{5pt}
  \subfloat[Mode 2 (pre-PIAA tilt) of the LOWFS.]
	{\label{fig:mode2}\FIG{0.235}{false}{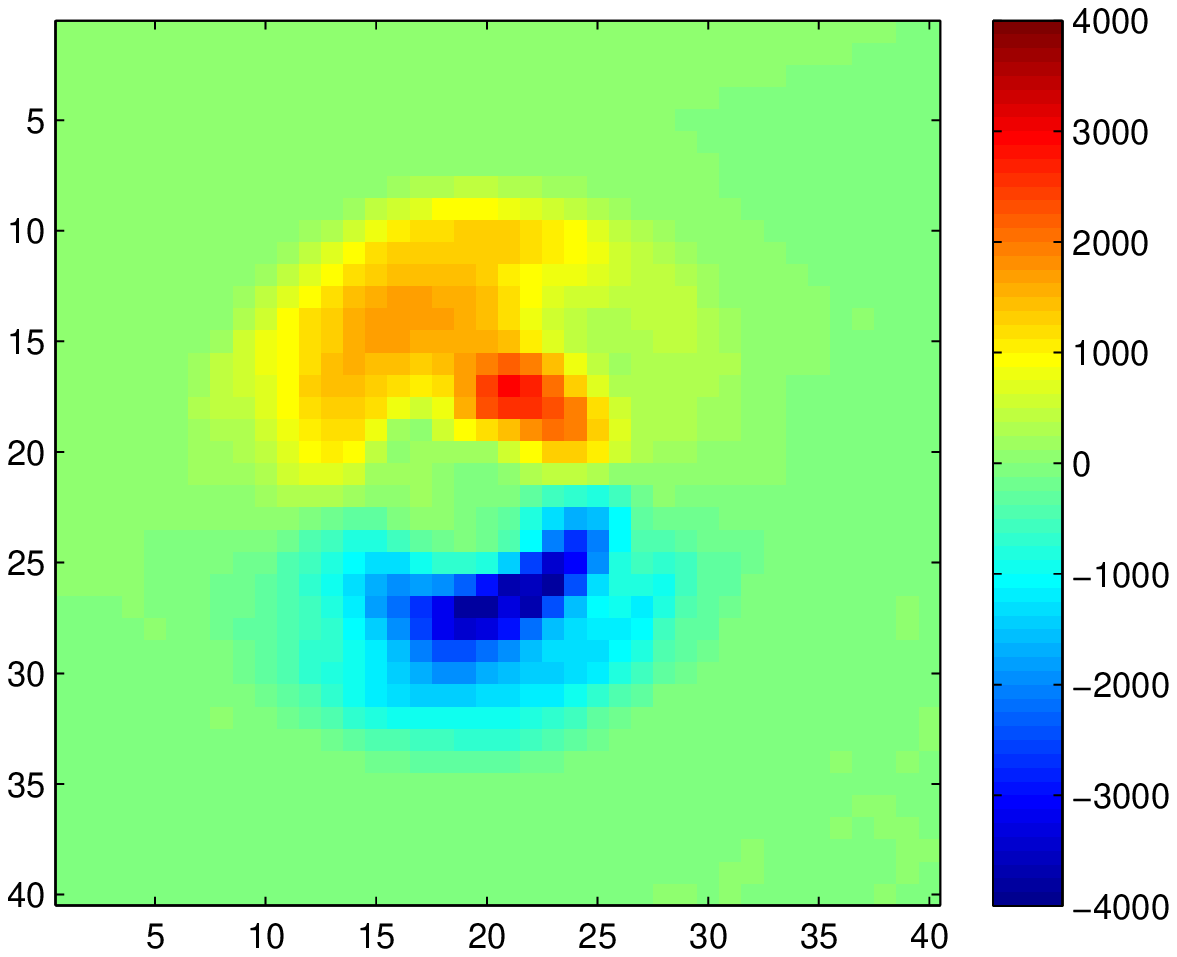}}
  \caption{Reference image and signal measured by the LOWFS.}
  \label{fig:images}
\end{figure}

Figure~\ref{fig:refim} presents a typical reference image, obtained by saving an average of a few hundred images after putting the coronagraphic mask at the desired position. The image is slightly defocused, and we can see the opaque disk at the center of the PSF. We defined the inner working angle (IWA) as the off-axis position of the source where the throughput is exactly 50\%. However, he size of the mask does not exactly match this definition, so it is usually not exactly centered on the PSF, we observe then an asymmetry.

After obtaining this reference image, it is subtracted to every images coming from the camera. An example of the resulting difference is shown in Fig.~\ref{fig:signal}.

For the second step, we send known voltages to the different actuators within their range of linearity. Then we measure the impact of each displacement on the signal. Figures~\ref{fig:mode1} and \ref{fig:mode2} present the result of the calibration of the modes. This calibration is similar to the construction of the influence matrix in adaptive optics. The measurement provided by the LOWFS is then the deconstruction of the signal presented in Fig.~\ref{fig:signal} into the modes presented in Fig~\ref{fig:mode1} and \ref{fig:mode2}.

The LOWFS we developed provides also a classical measurement of the centroid of the image. The centroid is then compared to the centroid of the reference image.


\subsection{Linearity}
\label{sec:Linearity}

\begin{figure} \centering
	\FIG{0.7}{false}{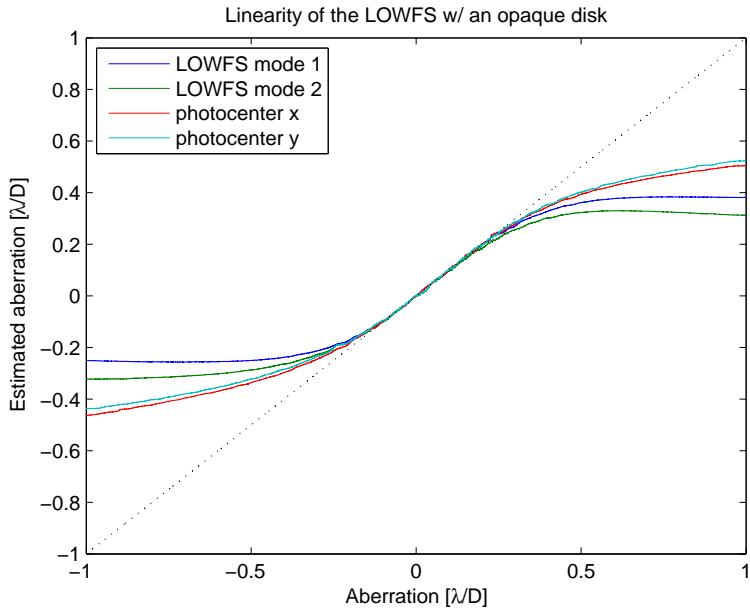}
  \caption{Linearity of the LOWFS with the 3-zone mask.}
  \label{fig:calib}
\end{figure}

To measure the linearity of both the LOWFS and the centroid measurement, we scanned between $-1$ and 1~\loD on both axes with the actuators. Figure~\ref{fig:calib} presents the experimental result. On this figure, the linearity range is approx. the same for both the LOWFS and the centroid measurement, about 0.4~\loD. For this mask, the centroid is monotonic over the whole range, while the LOWFS measurement has a plateau after $\pm0.5$~\loD.  We observe also a slight difference between modes 1 and 2 of the LOWFS, it is probably because the star is not perfectly centered on the mask.

So if we correct only pre-PIAA tip/tilt modes, the 3-zone mask provides a small linearity range, but it has the great advantage of reducing non-common path aberrations. Indeed, the LOWFS measurement is also sensitive to the position of the opaque disk on the star, while we observed non-common path aberrations with a reflective mask without opaque disk.


\section{Experimental results}
\label{sec:ExperimentalResults}


\subsection{Performances of the LOWFS}
\label{sec:PerformancesOfTheLOWFS}

For this first version of the LOWFS, the control algorithm is a simple integrator law, with a sampling frequency of 1.1~kHz. The command is delayed by two frames compared to the acquisition, so the rejection function has an overshoot around $f_\text{samp}/10$, with an amplitude dependent of the gain. So we have to use a small gain of 0.1, otherwise mechanical vibrations of the bench around 100~Hz are amplified.

As explained in \cite{Thomas13b}, the bench is covered with an enclosure, and the temperature is stabilized at a millikelvin level. Nevertheless, a part of the disturbance we observe is due to residual turbulence inside the enclosure. The rest is due to mechanical vibrations, coming from the bench and the different mounts. Then in open-loop, the tip/tilt disturbance is already low: $5.8\E{-3}$~\loD on the $x$-axis and $9.4\E{-3}$~\loD on the $y$-axis.

With a low gain of 0.1, the cutoff frequency of the integrator law is around $f_\text{samp}/100$, i.e. around 10~Hz. So we are not correcting most of the vibrations, but mostly the turbulence, which is below 1~Hz.

\begin{figure} \centering
  \subfloat[PSD of the LOWFS residue in closed-loop.]
	{\label{fig:vibpsd}\FIG{0.5}{false}{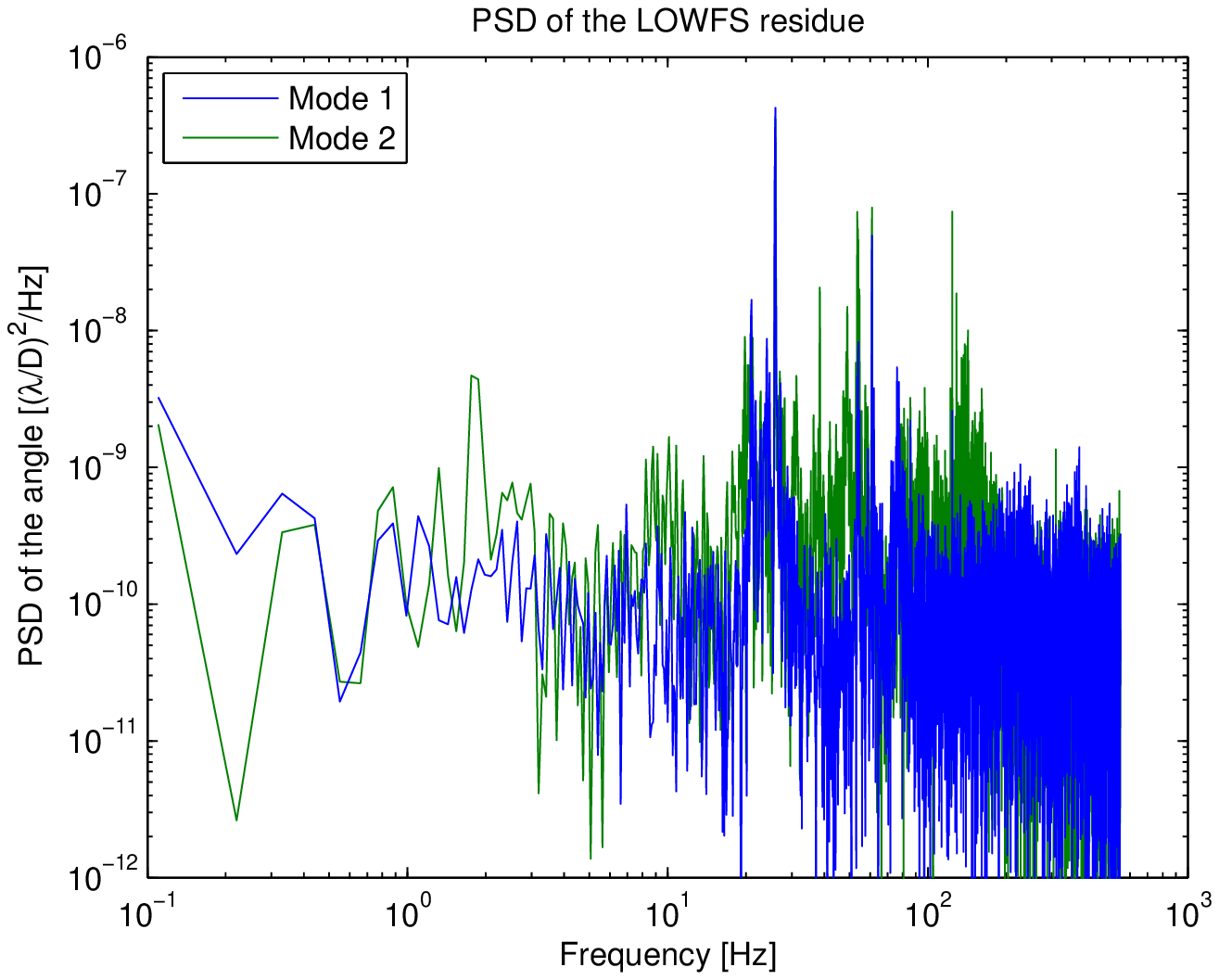}}
  \subfloat[Cumulative integral of the PSD of the residue.]
	{\label{fig:vibint}\FIG{0.5}{false}{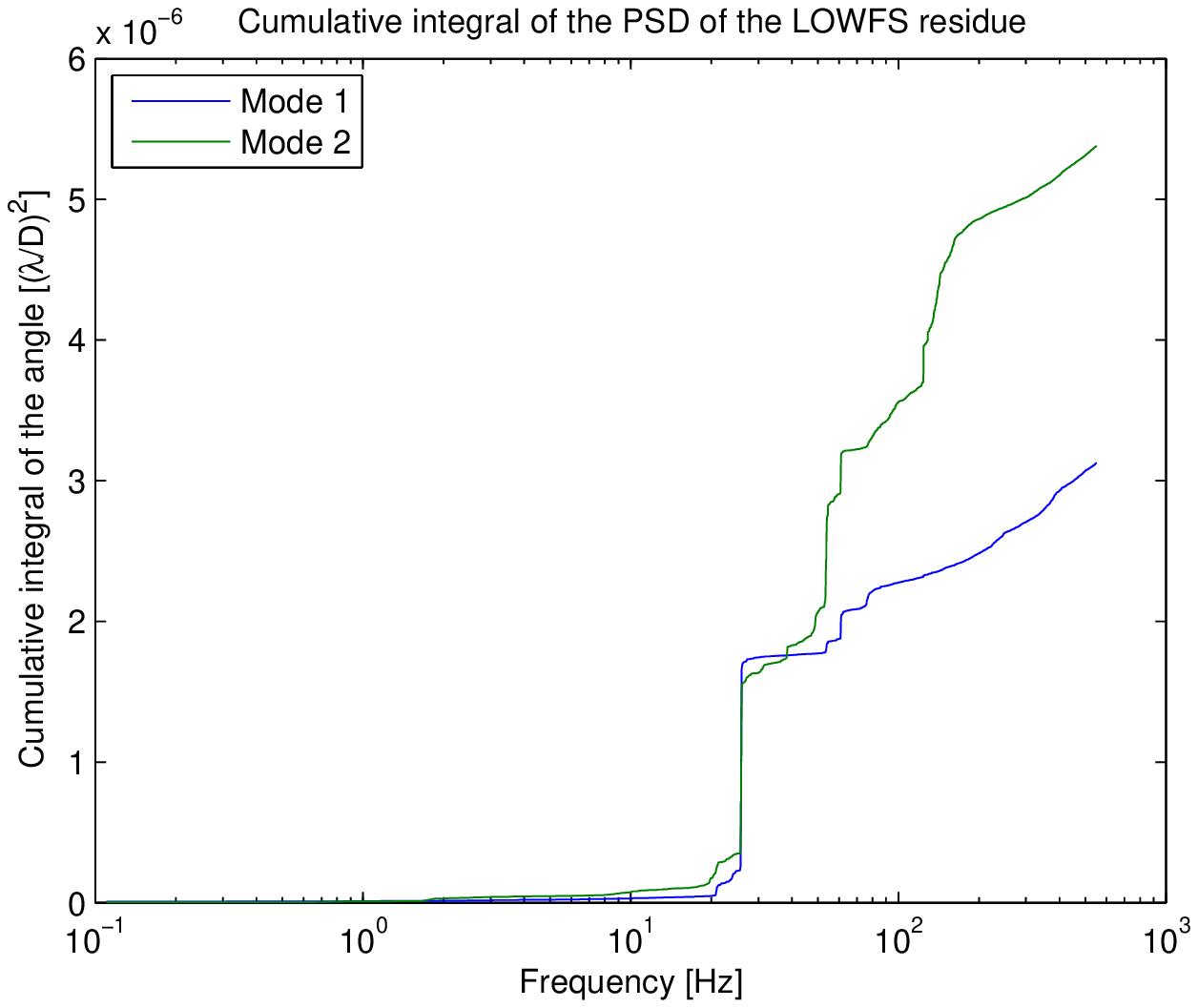}}
  \caption{Vibration analysis with the LOWFS.}
  \label{fig:vibration}
\end{figure}

Figure~\ref{fig:vibpsd} presents the Power Spectrum Density (PSD) of a 10~s measurement of the residue, and Fig.~\ref{fig:vibint} is the cumulative integral of that same PSD. In closed-loop, the residue drops to $1.5\E{-3}$~\loD on the $x$-axis, and $2.0\E{-3}$~\loD on the $y$-axis.

In Fig.~\ref{fig:vibpsd}, we can see that the residue is dominated by mechanical vibrations. The cumulative integral presented in Fig.~\ref{fig:vibint} is very useful to identify the strongest vibrations. Indeed, on this plot, vibrations appear as a sharp rise in the signal, while the maximum value in the plot corresponds to the variance of the residue. In this figure, we can see that the strongest vibration is at 25~Hz, contributing to half of the residue in $x$, and a third in $y$. On the $y$-axis, the rest comes mostly from a vibration at 53~Hz, and a wide contribution between 120 and 150~Hz.

To improve the correction, we planned to implement a Linear Quadratic Gaussian controller (LQG) with an automatic identification of the vibrations \cite{Meimon10}. It has already been successfully implemented on the \coro SPHERE \cite{Petit11} and on other experiments \cite{Petit08,Costille11,Lozi11}. With such a controller, we should be able to remove most of the vibrations. Simulations shows an improvement of about $5\E{-4}$~\loD on both axis.


\subsection{Contrast sensitivity to tip/tilt errors}
\label{sec:SensitivityOfTheContrast}

To achieve the desired contrast of $\e{-6}$ in the inner region between 1.2 and 2~\loD and $\e{-7}$ in the outer region between 2 and 4~\loD, we tested two different algorithms: Electric Field Conjugation (EFC) and speckle nulling \cite{Thomas13b}. Our current performance is better than the specifications: $1.8\E{-7}$ between 1.12 and 2~\loD, and $6.5\E{-8}$ between 2 and 4~\loD.

To understand if tip/tilt errors are limiting us on the bench, we injected different levels of tip/tilt errors: a random white noise with an adjustable amplitude, sent on the two piezo actuators controlling the position of the fiber. We measured each time the real amplitude of the noise, because it depends also on the disturbances already present on the bench, the response of the actuators, which modifies the amplitude of the noise at high frequency, and the response of the controller, which reduces the injected noise at low frequencies.

\begin{figure} \centering
\FIG{0.7}{false}{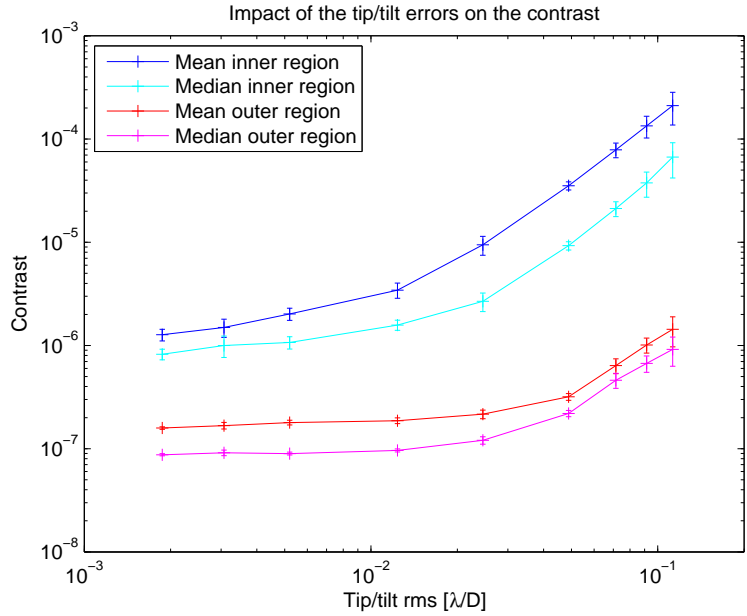}
\caption{Contrast at different levels of tip/tilt errors.}
\label{fig:contvstt}
\end{figure}

Figure~\ref{fig:contvstt} presents the results of that experiment. The initial contrast without injected noise was better than our goal, but significantly higher than our best performance: $8\E{-7}$ in the inner region, and $9\E{-8}$ in the outer region.

In this figure we can see two regimes:
\begin{itemize}
	\item The contrast is almost not affected by tip/tilt errors below $\e{-2}$~\loD rms in the inner region, and $3\E{-2}$~\loD rms in the outer region.
	\item The contrast is proportional to the square of the tip/tilt after $2\E{-2}$~\loD rms in the inner region, and around $5\E{-2}$~\loD rms in the outer region.
\end{itemize}

Thanks to this experiment, we deduced that we can accept tip/tilt errors up to almost $\e{-2}$~\loD rms and still meet our goal of a $\e{-6}$ contrast in the inner region.


\subsection{Impact of the wavefront control}
\label{sec:ImpactOfTheWavefrontControl}

Between iterations of the wavefront control, the shape of the PSF changes a lot, because the light inside the dark zone is shifted to the other side of the PSF. The LOWFS is sensitive to this effect, but it is then interpreted as a variation of tip/tilt.

\begin{figure}[b] \centering
  \subfloat[Profile of the PSF during the first seven iterations of EFC.]
	{\label{fig:psfwwfc}\FIG{0.485}{false}{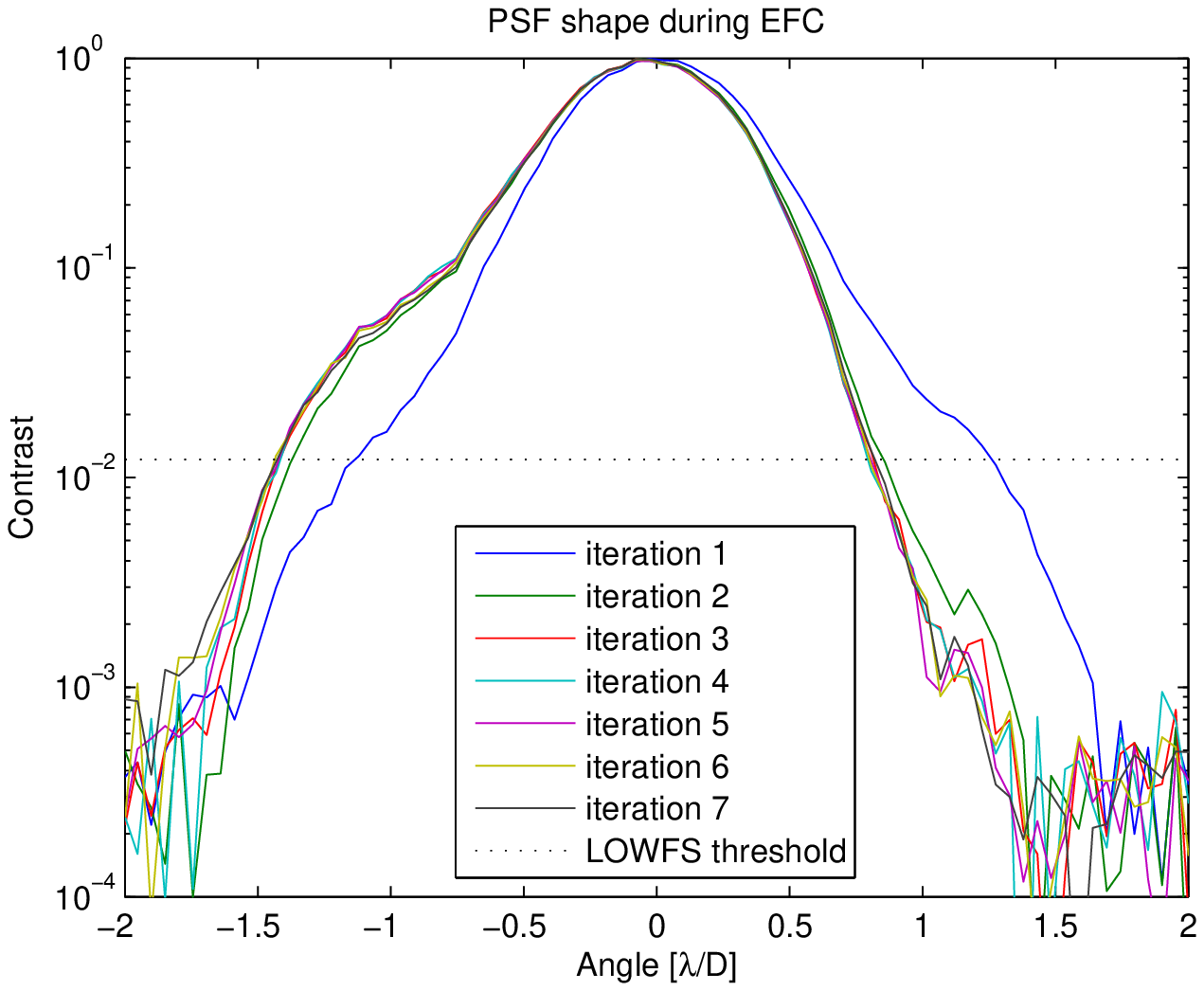}}\hspace{10pt}
  \subfloat[Measurement of the LOWFS in open loop during EFC.]
	{\label{fig:lowfswwfc}\FIG{0.485}{false}{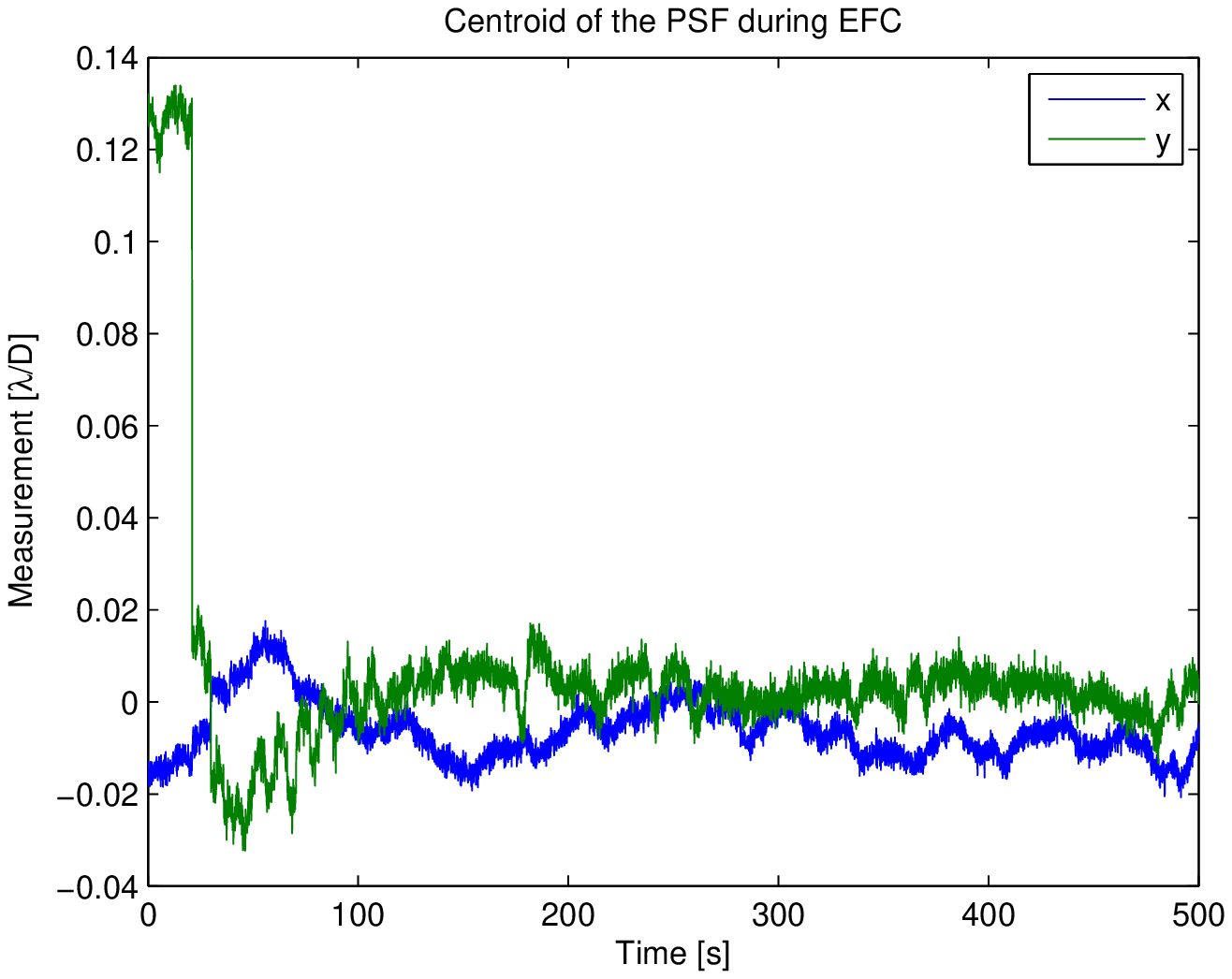}}
  \caption{Impact of the wavefront control on the PSF and the LOWFS.}
  \label{fig:wfc}
\end{figure}

Figure~\ref{fig:psfwwfc} presents the PSF profiles during the first seven iterations of EFC, taken with the science camera. Those profiles are along the $y$-axis of the bench, in the middle of the dark zone. On the first iteration, the DM is flat, and no dark zone is created. The PSF is then almost symmetric, with a contrast higher than $\e{-2}$ at 1.2~\loD.

After the first shape is sent to the DM, the contrast in the dark zone drops lower than $\e{-3}$, and after the third iteration, the modifications of shapes are below the readout noise of the science camera. In Fig.~\ref{fig:psfwwfc}, the dotted line corresponds to the noise level of the LOWFS images. We can see that after iteration~3, the shape of the PSF over the threshold does not change anymore.

This is confirmed by Fig.~\ref{fig:lowfswwfc}, corresponding to the LOWFS measurement during EFC. on the $y$-axis, we can see a first step at around 20~s, corresponding to the difference between iterations~1 and~2, and a second smaller step between iterations~2 and~3 at around 30~s. After those steps, the variations we observe correspond to other contributions, i.e. turbulence and vibrations. So the LOWFS is not affected by PSF changes after iteration~3.

The difference between a flat DM and a dark zone shifts the measurement of about 0.15~\loD, which corresponds to the correction we have to apply on the focal plane mask to keep the IWA at 1.2~\loD.

So the LOWFS should be used only after a few iterations of the WFC, to avoid any residual tip/tilt due to the correction. In any case, in the first iterations, the contrast is high enough that the system is not sensitive to tip/tilt errors, it is only useful when the contrast reaches at least $\e{-6}$ at 1.2~\loD.


\section{Application to the ELTs}
\label{sec:ApplicationToTheELTs}

Even though it is currently developed for a space mission, this type of LOWFS is well suitable for ground telescope, especially ELTs. With mirrors of that size, future coronagraphs will be extremely sensitive to tip/tilt disturbances, and if we want to obtain contrasts better than $\e{-7}$ at 2~\loD, we need to be able to control them at a microarcsecond level.

Even if the ELTs are much larger than the EXCEDE mission, and the space conditions are different from the ground, the layout of the LOWFS would actually be the same. But for ELTs, tip/tilt disturbances are much higher. We saw in Sec.~\ref{sec:Linearity} that the linearity range is fairly small, around 0.4~\loD. So this system would have to work in combination with at least one more stage of correction, for example the tip/tilt correction of the adaptive optic system. Vibrations will be a major issue, especially if their amplitude is higher than 0.4~\loD. But a solution like the LQG evoked in Sec.~\ref{sec:PerformancesOfTheLOWFS} should be able to manage them at the required level. 


\section{Conclusion}
\label{sec:Conclusion}

This paper presented a successful implementation of a coronagraphic low-order wavefront sensor. With a simple design, we were able to reduce pre-PIAA tip/tilt motions from between 5 and $9\E{-3}$~\loD rms in open-loop to $\approx2\E{-3}$~\loD rms in closed-loop. With a frame rate of 1.1~kHz, it was also very useful to perform a vibration analysis, leading to a modification of the most noisy elements. We also demonstrated that for the level of contrast and inner working angle we want for the EXCEDE mission, tip/tilt errors are currently not the limiting factor for our results.

This LOWFS is very interesting for any type of \coro using a focal plane mask, because its design is very simple, and it can be rapidly implemented on any testbench or instrument. It will be an asset for the \coros on the future ELTs, because it allows to correct tip/tilt and other low-order modes to a nanometric level, without introducing non-common path aberrations.

\end{document}